\documentclass[usenatbib,usegraphicx]{mn2e} 
\def\msun{M$_{\sun}$}  \def\aap{A\&A} \def\apjl{ApJ}
\def\apj{ApJ} \def\apjs{ApJS} \def\aj{AJ} \def\mnras{MNRAS} \def\nat{Nature}
\def\pasp{PASP}   \def\araa{ARA\&A}
\usepackage[flushleft]{threeparttable}
\usepackage{amsmath}

\title[The DECAM Minute Cadence Survey II]{The DECam Minute Cadence Survey II:
49 Variables but No Planetary Transits of a White Dwarf}

\author[K. Dame et al.] 
{Kyra Dame$^{1}$, 
Claudia Belardi$^{2}$, 
Mukremin Kilic$^{1}$, 
Armin Rest$^{3}$,
A. Gianninas$^{1}$, 
\newauthor Sara Barber$^{4}$,
Warren R. Brown$^{5}$\\ 
$^1$Homer L. Dodge Department of Physics \& Astronomy, University of Oklahoma, 440 W. Brooks St, Norman, OK 73019, USA\\
$^2$Department of Physics \& Astronomy, University of Leicester, University Road, Leicester LE1 7RH, UK\\ 
$^3$Department of Physics \& Astronomy, The Johns Hopkins University, 3400 North Charles Street, Baltimore, MD 21218, USA\\
$^4$House Committee on Science, Space, and Technology, 394 Ford House Office Building, Washington, DC 20515, USA\\ 
$^5$Smithsonian Astrophysical Observatory, 60 Garden St, Cambridge, MA 02138, USA\\ 
}

\begin{document}

\maketitle

\begin{abstract}

We present minute cadence photometry of 31732 point sources observed in one 3 $\rm deg^{2}$ DECam pointing 
centred at RA = 09:03:02 and Dec = -04:35:00 over eight consecutive half-nights. We use these data to search 
for eclipse-like events consistent with a planetary transit of a white dwarf and other sources of stellar 
variability within the field. We do not find any significant evidence for minute-long transits around our 
targets, hence we rule out planetary transits around $\sim 370$ white dwarfs that should be present in this 
field. Additionally, we identify 49 variables, including 40 new systems. These include 23 detached or contact 
stellar binaries, 1 eclipsing white dwarf + M dwarf binary, 16 $\delta$ Scuti, three RR Lyrae, and two ZZ Ceti 
pulsators. Results from the remaining two fields in our survey will allow us to place more stringent constraints 
on the frequency of planets orbiting white dwarfs in the habitable zone.
\end{abstract}


\section{Introduction}

Stellar variability on short timescales has been largely unexplored until recently,
but is relevant in the era of transient surveys like the Zwicky Transient
Facility \citep[ZFT,][]{bellm17}, the Large Synoptic Survey Telescope \citep[LSST,][]{ivezic08}, and in searches for optical counterparts
to gravitational wave events detected by the Laser Interferometer Gravitational-Wave Observatory
({\em LIGO}). Exoplanet surveys like the Wide Angle Search for Planets \citep[WASP,][]{pollacco06} and the
Hungarian-made Automated Telescope Network \citep[HATNet,][]{bakos04} have provided $\sim10$ minute cadence
observations of relatively bright stars, which can be used to search for transits around compact objects
like white dwarfs \citep{faedi11}. On the fainter end, the Sloan Digital Sky Survey Stripe 82
\citep{bramich08}, the Panoramic Survey Telescope \& Rapid Response System (Pan-STARRS) Medium Deep Fields
\citep{tonry12}, and the Palomar Transient Factory (PTF, \citealt{rau09}) provided nightly
observations of large areas of sky that were specifically designed to detect supernovae.
\citet{fulton14} used the Pan-STARRS data on $\sim1700$ photometrically selected candidates to constrain
the frequency of gas giant planets orbiting white dwarfs just outside the Roche limit (0.01 AU) to $<0.5$\%, but
their cadence and sensitivity were not high enough to constrain the frequency of Earth-size planets at the same orbital
separation. 

The Kepler/K2 mission has provided short-cadence, 1 min, data for hundreds of objects, but mainly
for previously known pulsating stars. K2 long-cadence data led to
the discovery of a disintegrating asteroid around the dusty white dwarf WD 1145+017
\citep{vanderburg15,gansicke16,rappaport16}. Such planetesimals had never been
found around white dwarfs previously due to a lack of extended duration
high-cadence observations of a large number of white dwarfs. \citet{vansluijs18} used long and short cadence
K2 data from 1148 confirmed and high-probabality white dwarfs to constrain the frequency of Earth-sized planets 
in the habitable zone around white dwarfs to $<28$\%. For white dwarfs with masses between 0.4 and 0.9 \msun and 
$T_{\rm eff} < 10000$ K, the continuously habitable zone (the region where a planet could sustain liquid water 
for $>$ 3 Gyrs) extends from $\approx$ 0.005 to 0.02 AU \citep{agol11}.

The OmegaWhite Survey \citep{macfarlane15,macfarlane17,toma16} is observing 400 $\rm deg^{2}$ of
the sky along the Galactic Plane with a median cadence of 2.7 min, with the goal
of finding short period ($P_{orb}$ $<$ 80 min) variables, especially
ultracompact binaries as these are predicted to be the strongest emitters
of gravitational waves in the milli-Hertz frequency range \citep{amaro13,roelofs07}. 
However, the 2 hour duration of their observations is not ideal
for finding transits around white dwarfs as planets with such short periods would be
tidally disrupted \citep{agol11}.

The DECam Minute Cadence Survey \citep{belardi16} was the first high cadence survey of its
kind, observing 9 $\rm deg^{2}$ of the sky at a cadence of $\approx$ 90 sec over 8 half-nights per field,
specifically looking for transits around a large number of white dwarfs.
Initial results from this survey focused on 111 high proper motion white dwarfs in their
first 3 $\rm deg^{2}$ field. While \citet{belardi16} did not find evidence of planetary mass
companions to these high proper motion white dwarfs, they note that there should be hundreds
of white dwarfs with low proper motion ($\mu$ $<$ 20 mas $yr^{-1}$) in the same field. Here we extract
and analyse the photometry for all of the point sources in this field to search for
transiting and variable objects, including the lower proper motion objects.

We discuss the details of our observations and reductions in Section
\ref{sec:obs}. We discuss the results of our transit search in Section
\ref{sec:transits}, present our selection process and list of variable objects in Section
\ref{sec:variability}, and conclude in Section \ref{sec:con}.

\section{Observations and Data Reduction} \label{sec:obs}

\subsection{DECam Observations}

We used the Cerro Tololo Interamerican Observatory 4m telescope equipped with
the Dark Energy Camera (DECam) over eight half-nights in Feb 2014. We obtained $g$-band
exposures of a three square degree field, previously observed by the
Canada-France-Hawaii Telescope Legacy Survey \citep[CFHTLS,][]{cuillandre12} with
Megacam $ugriz$ photometry available. Our DECam field covers a small portion of
the CFHTLS Wide 2 Field. Exposure times were chosen to obtain S/N
$\geq$ 5 photometry of targets brighter than $g = 24.5$ AB mag, giving an
overall cadence of $\approx$ 90 s, including the 20 s read-out time of the camera.
Further details of the observations can be found in \citet{belardi16}. 

\subsection{Photometric Data Reduction}

We downloaded the raw DECam data from the NOAO Archive and used the Photpipe pipeline,
which was previously used in time-domain surveys like SuperMACHO, ESSENCE, and Pan-STARR1
\citep{rest05,garg07,miknaitis07,rest14}, to reduce these data. The pipeline performs single-epoch
image calibration, including bias subtraction, cross-talk correction,
flat-fielding, astrometric calibration, and geometrical distortion correction \citep[using the SWarp 
software package,][]{bertin02}. It performs photometry using a modified version of DoPHOT 
\citep{schechter93}, and applies quality cuts to the resultant catalogue to remove
stars less than 20 pixels from the edges of the CCDs, stars with a significantly brighter
neighbour, and stars with photometric errors greater than 3$\sigma$ (where
$\sigma$ is the mean photometric error of all objects in the corresponding 0.5
mag bin). Given our interest in variable objects, only relative photometry is
needed, and we do not perform absolute photometric calibration.

\subsection{Light Curve Creation}

In order to remove the effects of short-term changes in the atmosphere (cloud
coverage, haze, etc) and changes in airmass, we select ten bright, unsaturated,
non-variable stars from each CCD to use as reference stars. For each CCD, we
shift the light curves of our ten reference stars from that CCD to the same
magnitude scale and apply a sigma-clipping algorithm to remove bad points affected by
cosmic rays or CCD defects. We create a single reference light curve from the
weighted means of the individual light curves. Sigma clipping is only used in 
the creation of our calibration light curve and is not applied to our sources in 
any subsequent steps. We then subtract this calibration light curve from every source 
identified in the corresponding CCD. The error introduced by our calibration procedure 
is two orders of magnitude smaller than the overall scatter in the light curves. Note that this process 
was run separately for each night.

Given that our reference stars are typically redder than white dwarfs,
airmass-related effects are still visible in the light curves of many of our
sources; these effects lead to significant peaks in the Fourier Transform,
especially around 4 cycles per day (see Section \ref{sec:var}). To remove this
effect, we additionally fit a third-order polynomial to the calibrated source
light curves.

\subsection{Optical Spectroscopy}

We obtained follow-up optical spectroscopy of two of the variable sources using
the Blue Channel Spectrograph on the 6.5m MMT in October 2017. We operated the
spectrograph with the 832 line mm$^{-1}$ grating in second order, providing
wavelength coverage from 3600 \AA\ to 4500 \AA\ and a spectral resolution of 1.0 \AA.
We obtained all observations at the parallactic angle, with a comparison
lamp exposure paired with every observation. We flux-calibrated using blue
spectrophotometric standards \citep{massey88}.

We obtained follow-up optical spectroscopy of two additional variable sources using
the Gemini Multi-Object Spectrograph on the 8m Gemini South telescope as part of the 
programme GS-2018A-Q-319. Observations were obtained using the B600 grating and a 1.0
arcsec slit, providing wavelength coverage from 3650 \AA\ to 6750 \AA\ and a spectral 
resolution of 3.6 \AA. Targets were flux-calibrated using the spectrophotometric standard 
LTT7379. 

\section{Transits} 
\label{sec:transits}

Planetary eclipses around white dwarfs should only last a few min, hence such events 
would only affect one or two data points per orbital period. Given the extremely short
durations, traditional eclipse search algorithms like the box-least-squares periodogram 
are not a good way to identify white dwarf transits. Instead, a simple search for $5\sigma$
significant dips in flux is sufficient. However, identifying the source of these drop-outs, whether
they are intrinsic to the source or not (due to instrumental or sky background problems), can be
difficult. Using a $5\sigma$ threshold, \citet{fulton14} find 11,570 drop-outs (0.27\%)
out of 4.3 million data points in the Pan-STARRS Medium Deep Field data. However, further
inspection of the images with the drop-out points show that none are compatible with eclipses
by substellar companions to white dwarfs in those fields. 

\begin{figure}
\includegraphics[width = 10cm]{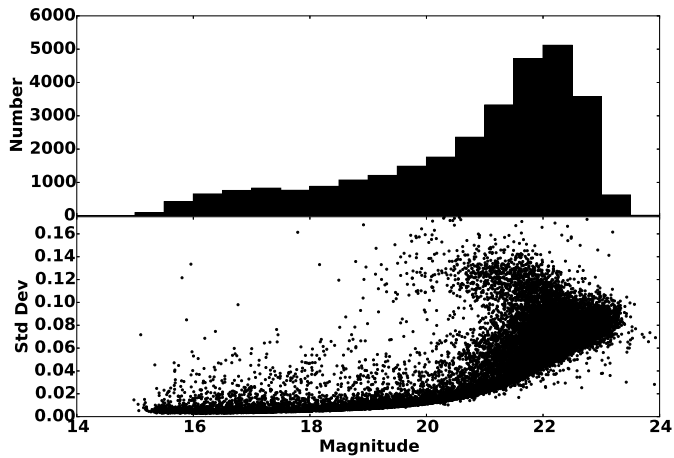}
\caption{Magnitude distribution (top) and plot of standard deviation (bottom) of the light curve as a function of magnitude for
 the point sources included in our analysis. The hook feature seen in the standard deviation plot around \textit{g}=22 mag is due to
 contamination from barely resolved sources.}
\label{fig:bin}
\end{figure}

Figure \ref{fig:bin} shows the magnitude distribution of the point sources in our first DECam field
(top), as well as the standard deviation $\sigma$ of the light curve of each source as a function of 
magnitude (bottom). This figure shows that our sample is incomplete beyond $g=22.5$ mag. The hook feature seen 
around \textit{g}=22 mag in the standard deviation plot is due to contamination from barely resolved 
sources in our sample. Even though the limiting magnitude of the individual DECam images is 
$g\approx$24 mag, our data reduction pipeline only includes sources that are detected at a signal-to-noise 
ratio S/N$\geq$10. This S/N ratio enables us to identify eclipses as shallow as 0.5 mag at $5\sigma$ 
confidence level, which is essential for finding Earth-size planets.

There are 31732 detections with point spread functions consistent with point sources in our DECam field.
We have a total of 1024 images, which results in more than 30 million data points.
We check each light curve for significant ($\geq$ 5$\sigma$)
dips and visually inspect the images with potential transits. We find 5244
potential transits, a significantly smaller fraction compared to the \citet{fulton14} study. However,
our imaging pipeline already removes most sources near the edges and near bright, saturated stars. Visual inspection 
of these dips shows that the majority of these sources are either close to the edge
of the chip, near bad pixels, or elongated, and are thus also not real. The point spread function
is worse near the edges of the DECam field of view and it gets worse at high airmass (toward the
end of the night). This explains many of the elongated images, which result in a lower
flux measurement in our photometry.

\begin{figure}
\includegraphics[width = 9cm]{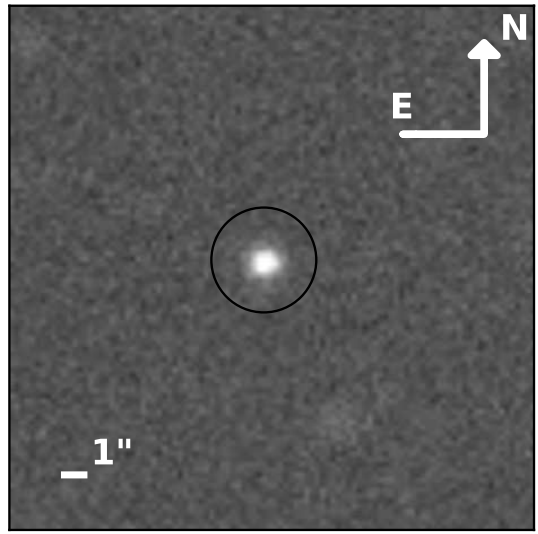}
\includegraphics[width = 9cm]{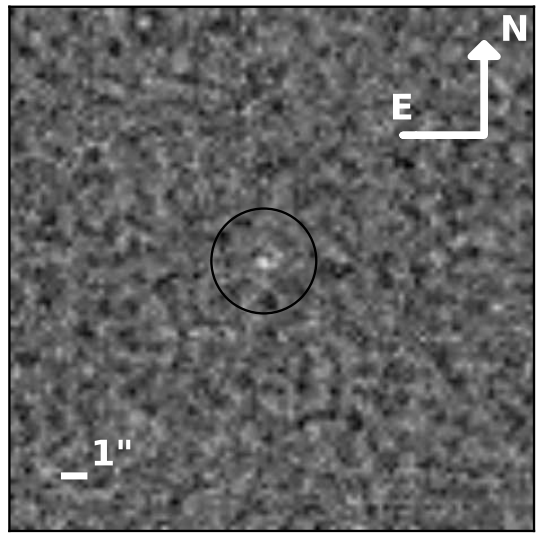}
\caption{Image taken on UT 2014 Feb 05 at 05:21:18, corresponding to the drop-out candidate at RA = 09:07:10 and Dec = -04:29:44 (top) 
and the difference image (bottom). The target is circled.}
\label{fig:image}
\end{figure}

Of our 5244 potential transits, we find eight that appear to be real, though interestingly, no source exhibits 
more than one genuine drop-out. However, even with a 5$\sigma$ threshhold and assuming gaussian errors, we still 
expect $\approx$18 such deviations in a sample of 30 million points simply due to random scatter. Hence, we expect 
to find about nine $5\sigma$ drop-outs. To determine whether these eight drop-outs are real, we visually inspect 
difference images obtained with High Order Transform of PSF and Template Subtraction \citep[HOTPANTS,][]{becker15}. 
Figure \ref{fig:image} shows the image and difference image of an example drop-out candidate.
We observe no corresponding dip in the difference images for our drop-out candidate, 
although it should be noted that for five of our eight drop-outs, the star is poorly subtracted. Therefore, while we do find 
eight dips consistent with a planetary transit of a white dwarf in our photometry, none of these dips appear 
significant to 3$\sigma$ in the difference images so they are likely not real. 

Of our 31732 targets, only 5011 are detected in \textit{GAIA} Data Release 2 \citep{gaia18} with parallax over error $\geq$ 4, 
and so \textit{GAIA} is not useful for estimating the total number of white dwarfs in our field at these faint magnitudes. 
However, the Besan{\c c}on Galaxy model \citep{robin03} predicts 30381 stars with 13 $\leq \textit{g} \leq$ 24 in our field, 
in good agreement with our 31732 detections. The model also predicts ~374 white dwarfs within our field, giving  a WD fraction 
of 1.23\%. \citet{belardi16} calculated an expected detection rate of 0.7\% due to the observing window from the ground for 
orbital periods less than 30 h. Therefore, based on these numbers, we would expect to find 2.6 planets in our sample if every 
white dwarf had an earth-mass planetary companion in its habitable zone. 

\section{Variability}
\label{sec:variability}

\subsection{Variable Candidate Selection} \label{sec:var}

\begin{figure}
\hspace{-0.2in}
\includegraphics[width = 10cm]{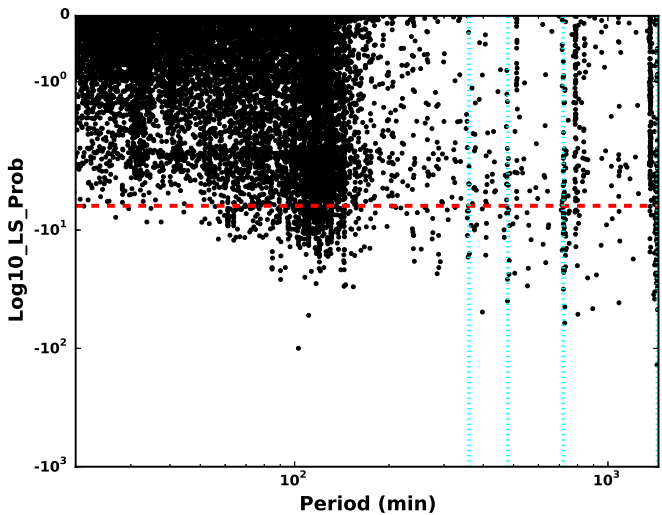}
\caption{The false alarm probability ($log_{10}(FAP)$) as a
 function of period for our sample. The red dashed line shows the theoretical
 5$\sigma$ confidence level; we select all objects below this line as our
 initial variable candidates. Cyan dotted lines mark periods of 6, 8, 12, and 24 hrs.}
\label{fig:fap}
\end{figure}

Using the VARTOOLS suite of software \citep{hartman16}, we run a Lomb-Scargle
periodogram analysis on all the sources in the field. From this, we get the five
most significant peaks in the periodogram and their corresponding false alarm
probability (FAP). Figure \ref{fig:fap} shows the distribution of
$log_{10}$(FAP) values as a function of period for our sample. In theory, a star
with $log_{10}$(FAP) = $-6.2$ has a probability of being a variable object at the
5$\sigma$ confidence level, with values more negative having higher probability
of real variability. We use this limit to select our initial variable
candidates. However, as discussed in \citet{macfarlane15}, the 5$\sigma$
confidence level can be more negative for real data than the theoretical limit
due to systematic effects such as red noise, which introduces a larger spread in
$log_{10}$(FAP) values at longer periods.

Of our 31732 sources, we find 889 variable candidates with at least one period
with $log_{10}$(FAP) $\leq -6.2$. However, as mentioned above, this is 
only a theoretical limit, and we expect the real 5$\sigma$ confidence level to 
be more negative, especially at longer periods. Therefore, we use the Period04 
package \citep{lenz14} to inspect each candidate. We compare 
the amplitudes of the observed peaks with the median amplitude, $\langle$A$\rangle$, 
of the Fourier Transform and look for significant peaks ($\geq$ 5$\langle$A$\rangle$). 
Of our initial 889 candidates, 213 show no evidence of a significant peak upon inspection
with Period04. For the remaining targets, we phase-fold the light curves to the period 
determined by Period04. Many of the resulting phase-folded light curves exhibit obvious 
problems due to attempts to fit outliers and minor zero-point magnitude differences across nights 
(due to our calibration method). Additionally, we see a significant number of 
false positives, occurring around 6, 8, 12, and 24 hours (see Fig. \ref{fig:fap}) due to 
our observing window from the ground. These problems result in poor phase coverage,
and the light curves for these objects show no compelling evidence of variation at the 
periods and amplitudes determined.

However, we find 49 objects that do show genuine variability in their light curves. Table 
\ref{tab:var} presents the coordinates, periods, CFHT Legacy Survey $ugriz$ photometry, and
variable types for these sources. Note that for some, our method has difficulty finding an accurate 
period; most of these objects appear to have periods longer than our 4 hour nightly observing window, 
leading to poor phase coverage (see Section \ref{sec:binary}). However, their light curves
do exhibit clear variability. For example, J0902$-$0510, J0903$-$0516,
and J0905$-$0438 are eclipsing detached binary systems, but our phase coverage is not
good enough to recover the binary period.

\begin{figure}
\hspace{-0.3in}
\includegraphics[width = 10cm]{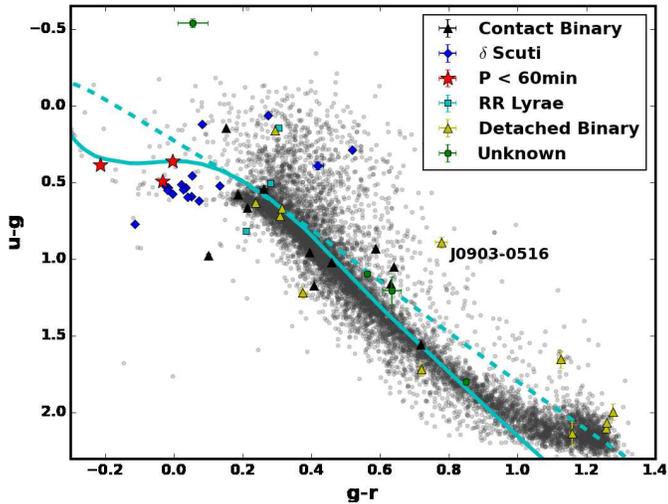}
\caption{Colour-colour diagrams for
 our 49 variable objects and non-variable objects with \textit{g}$<$20 mag (grey points). 
 Colors are taken from the CFHT Legacy Survey $ugriz$ photometry (AB scale). Predicted 
 colors for log g = 8.0 white dwarfs with pure hydrogen atmospheres (solid line) and pure helium 
 atmospheres (dashed line) are shown for reference. The white dwarf + M dwarf binary J0903$-$0516 
 is labelled.}
\label{fig:colour}
\end{figure}

For the majority of our targets, we are able to determine the type of variable star
from the colours, period, and general shape of the light curve. Figure \ref{fig:colour}
shows the colour-colour diagram for our variable targets, separated by type, along with
all non-variable objects with \textit{g}$<$20 in the field, and the predicted colors\footnote{http://www.astro.umontreal.ca/~bergeron/CoolingModels/} 
for pure hydrogen and pure helium white dwarfs \citep{holberg06,kowalski06,tremblay11,bergeron11}. 
The CFHT Legacy Survey Wide 2 Field has low-extinction, $E(B-V) = 0.02$ mag. Hence, most pulsating 
stars ($\delta$ Scuti and RR Lyrae) cluster around relatively blue colors
($g-r \sim 0$), and are easy to identify, whereas detached and contact binaries involve stars with
many different spectral types. Hence, these binaries appear everywhere along the stellar sequence.

\begin{figure}
\hspace{-0.3in}
\includegraphics[width = 10cm]{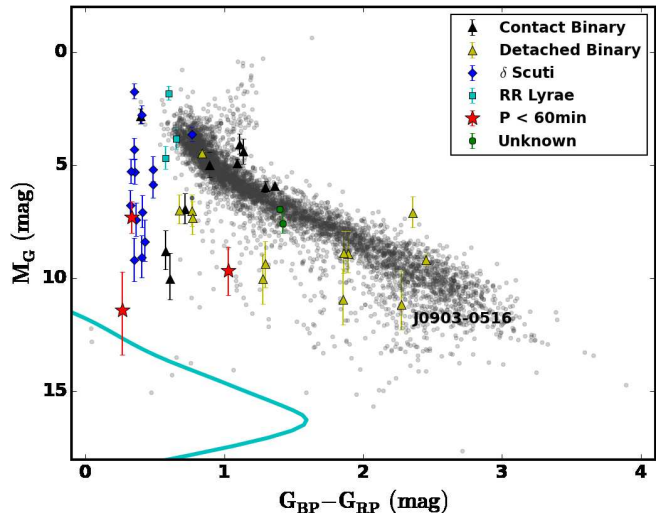}
\caption{Colour-magnitude diagram for
 our 45 variable objects detected with \textit{GAIA}. Symbols are the same as Figure \ref{fig:colour}. 
The 5001 non-variable sources from our survey with parallax over error $\geq$ 4 are shown as grey points. 
The cyan line shows the model sequence for a 0.6\msun H atmosphere white dwarf. The white dwarf + M dwarf 
binary J0903$-$0516 is labelled.}
\label{fig:cmd}
\end{figure}

Additionally, 45 of our variable sources were detected with \textit{GAIA}. However, 11 have 
negative parallax values and only 10 have parallax over error $\geq$ 4. We use the geometric distance
estimates from \citet{bailer18}, inferred from Bayesian analysis with a weak distance prior based on an 
exponentially decreasing space density such that the probability \textit{P(r|L)}

\begin{equation}
    P(r|L)=
\begin{cases}
    \frac{1}{2L^{3}}r^{2}e^{-r/L},& \text{if}\ r>0 \\
    0,                            & \text{otherwise}
\end{cases}
\end{equation}\\

where \textit{r} is the distance and $L>0$ is a scale length.

In Figure \ref{fig:cmd}, we show the colour-magnitude diagram for 
these 45 sources. We also include the 5001 non-variable sources (grey points) detected with \textit{GAIA} with 
parallax over error $\geq$ 4 and the model sequence (cyan line) for a 0.6\msun H atmosphere white dwarf. 
The main sequence is clearly visible, and the positions of our variable sources show good agreement with 
our classifications.

In the following sections, we discuss our results for each variable type.

\begin{figure}
\includegraphics[width = 10cm]{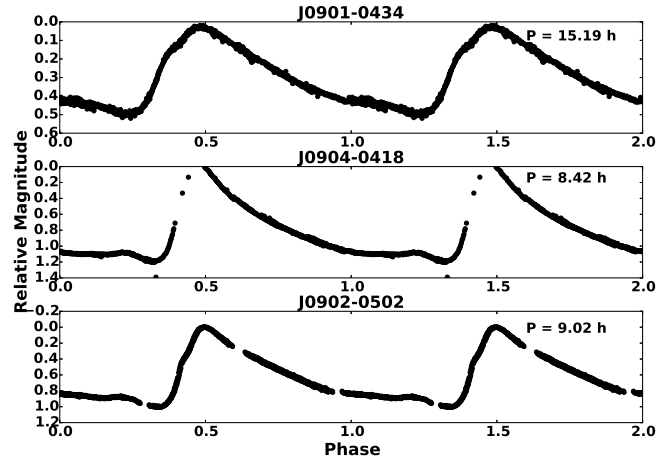} 
\caption{Phase-folded light curves for the three previously known
RR Lyrae variables.}
 \label{fig:rr}
\end{figure}

\begin{table*}
\centering
\scriptsize
\caption{Coordinates, periods, \textit{ugriz} photometry from the CFHT Legacy Survey, and determined variable type for out 49 variable candidates. Sources for previously known variables are provided.}
\begin{threeparttable}
\label{tab:var}
\begin{tabular}{lccccccccc}
\hline
Object & RA & Dec & Period &        u  &         g  &         r  &         i  &         z  & Type\\
       &    &     & (hr)   &           &            &            &            &            &     \\
\hline
$J0859-0439$  & 08:59:37.250 & $-$04:39:29.27 & 7.21  & 22.79 & 21.07 & 20.35 & 19.74 & 19.63 & Detached\\
$J0900-0426$  & 09:00:20.133 & $-$04:26:27.36 & 46.15 & 22.43 & 21.21 & 20.84 & 20.21 & 20.30 & Detached\\
$J0901-0350$ \tnote{c} & 09:01:03.319 & $-$03:50:42.54 & 6.80  & 16.96 & 15.91 & 15.27 & 14.94 & 14.39 & Contact\\
$J0901-0351$  & 09:01:23.589 & $-$03:51:06.45 & 21.82 & 19.66 & 18.94 & 18.63 & 18.51 & 18.52 & Detached\\
$J0901-0352$  & 09:01:29.915 & $-$03:52:34.44 & 7.19  & 21.76 & 21.22 & 20.95 & 20.92 & 20.90 & Contact\\
$J0901-0525$  & 09:01:38.071 & $-$05:25:05.08 & 96.00 & 19.16 & 18.53 & 18.29 & 18.21 & 18.48 & Detached\\
$J0901-0457$  & 09:01:56.909 & $-$04:57:04.07 & 6.96  & 19.99 & 19.33 & 19.11 & 18.96 & 18.92 & Contact\\
$J0902-0422$  & 09:02:21.571 & $-$04:22:42.30 & 60.00 & 22.78 & 20.79 & 19.51 & 18.65 & 18.41 & Detached\\
$J0902-0439$  & 09:02:45.974 & $-$04:39:34.64 & 7.08  & 21.04 & 20.07 & 19.97 & 19.83 & 19.64 & Contact\\
$J0902-0353$ \tnote{c} & 09:02:48.028 & $-$03:53:03.22 & 7.60  & 18.10 & 16.93 & 16.52 & 16.51 & 15.91 & Contact\\
$J0902-0510$  & 09:02:57.061 & $-$05:10:20.81 & \dots & 20.20 & 19.53 & 19.22 & 19.10 & 19.23 & Detached\\
$J0903-0516$  & 09:03:24.589 & $-$05:16:53.25 & \dots & 22.72 & 21.83 & 21.05 & 19.70 & 19.10 & Detached\\
$J0903-0429$  & 09:03:42.089 & $-$04:29:40.20 & 6.09  & 23.10 & 21.96 & 19.80 & 18.97 & 18.63 & Detached\\
$J0903-0448$  & 09:03:44.624 & $-$04:48:32.97 & 14.04 & 21.91 & 19.81 & 18.55 & 17.40 & 16.96 & Detached\\
$J0904-0415$ \tnote{c} & 09:04:07.252 & $-$04:15:53.13 & 6.08  & 18.82 & 17.26 & 16.55 & 16.75 & 16.18 & Contact\\
$J0904-0438$  & 09:04:15.549 & $-$04:38:03.73 & 6.22  & 18.80 & 17.64 & 17.01 & 17.07 & 16.62 & Contact\\
$J0904-0429$  & 09:04:31.582 & $-$04:29:29.13 & 8.89  & 17.51 & 16.58 & 16.00 & 16.33 & 15.53 & Contact\\    
$J0904-0516$ \tnote{c} & 09:04:32.882 & $-$05:16:57.97 & 10.17 & 15.97 & 15.82 & 15.67 & 15.64 & 15.65 & Contact\\
$J0904-0404$  & 09:04:37.052 & $-$04:04:23.43 & 5.97  & 20.11 & 19.53 & 19.34 & 19.33 & 19.34 & Contact\\
$J0904-0450$  & 09:04:42.776 & $-$04:50:55.14 & 7.29  & 20.57 & 19.55 & 19.09 & 19.11 & 19.11 & Contact\\
$J0905-0520$ \tnote{c} & 09:05:09.954 & $-$05:20:49.69 & 28.57 & 15.75 & 15.59 & 15.29 & 15.18 & 14.87 & Detached\\
$J0905-0438$  & 09:05:16.806 & $-$04:38:18.66 & \dots  & 21.82 & 19.75 & 18.50 & 17.49 & 16.83 & Detached\\
$J0905-0511$  & 09:05:49.293 & $-$05:11:46.41 & 26.09  & 23.18 & 21.53 & 20.40 & 19.75 & 19.55 & Detached\\
$J0905-0507$ \tnote{c} & 09:05:51.584 & $-$05:07:08.31 & 8.51  & 18.74 & 17.78 & 17.39 & 17.34 & 17.25 & Contact\\
\hline
\hline
Object & RA & Dec & Period &        u  &         g  &         r  &         i  &         z  & Type\\
       &    &     & (hr)   &           &            &            &            &            &\\
$J0901-0434$ \tnote{b} & 09:01:41.200 & $-$04:34:10.06 & 15.19 & 18.02 & 17.21 & 17.00 & 16.89 & 16.89 & RR Lyrae\\
$J0902-0502$ \tnote{b} & 09:02:22.283 & $-$05:02:01.24 & 9.02  & 17.98 & 17.48 & 17.20 & 17.36 & 17.49 & RR Lyrae\\
$J0904-0418$ \tnote{b} & 09:04:57.676 & $-$04:18:15.11 & 8.42  & 16.00 & 15.85 & 15.54 & 16.02 & 15.15 & RR Lyrae\\
\hline
\hline
Object & RA & Dec & Period &        u  &         g  &         r  &         i  &         z  & Type\\
       &    &     & (min)  &           &            &            &            &            &\\
$J0900-0444$  & 09:00:22.458 & $-$04:44:12.28 & 69.03  & 19.76 & 19.22 & 19.19 & 19.23 & 19.33 & $\delta$ Scuti\\
$J0900-0358$  & 09:00:33.589 & $-$03:58:24.21 & 91.95  & 18.72 & 18.10 & 18.03 & 18.03 & 18.07 & $\delta$ Scuti\\
$J0900-0456$  & 09:00:49.936 & $-$04:56:24.71 & 73.13  & 17.62 & 17.17 & 17.12 & 17.18 & 17.26 & $\delta$ Scuti\\
$J0901-0437$  & 09:01:53.875 & $-$04:37:14.94 & 96.6   & 21.93 & 21.54 & 21.12 & 21.06 & 20.97 & $\delta$ Scuti\\
$J0902-0351$  & 09:02:33.540 & $-$03:51:50.44 & 93.6   & 18.42 & 17.65 & 17.76 & 17.65 & 17.73 & $\delta$ Scuti\\
$J0903-0519$  & 09:03:16.452 & $-$05:19:11.81 & 55.41  & 20.55 & 20.01 & 20.03 & 20.08 & 20.16 & $\delta$ Scuti\\
$J0904-0506$  & 09:04:11.330 & $-$05:06:22.54 & 59.26  & 19.58 & 19.07 & 19.05 & 19.10 & 19.18 & $\delta$ Scuti\\
$J0904-0442$  & 09:04:29.256 & $-$04:42:50.98 & 67.51  & 19.30 & 18.71 & 18.66 & 18.74 & 18.83 & $\delta$ Scuti\\
$J0904-0423$  & 09:04:50.248 & $-$04:23:43.93 & 57.19  & 18.16 & 17.62 & 17.59 & 17.79 & 17.77 & $\delta$ Scuti\\
$J0904-0515$  & 09:04:50.834 & $-$05:15:34.55 & 91.78  & 16.25 & 16.13 & 16.04 & 16.08 & 16.08 & $\delta$ Scuti\\
$J0905-0437$  & 09:05:28.246 & $-$04:37:26.78 & 67.61  & 15.78 & 15.72 & 15.45 & 16.15 & 15.53 & $\delta$ Scuti\\
$J0905-0433$  & 09:05:35.177 & $-$04:33:54.28 & 56.03  & 19.24 & 18.72 & 18.74 & 18.86 & 18.98 & $\delta$ Scuti\\
$J0905-0430$  & 09:05:40.849 & $-$04:30:17.85 & 59.33  & 18.48 & 17.93 & 17.95 & 18.03 & 18.12 & $\delta$ Scuti\\
$J0905-0458$  & 09:05:45.608 & $-$04:58:27.99 & 59.90  & 20.20 & 19.68 & 19.54 & 19.77 & 19.86 & $\delta$ Scuti\\
$J0906-0407$  & 09:06:21.666 & $-$04:07:23.86 & 15.02  & 16.81 & 16.52 & 16.00 & 16.46 & 15.70 & $\delta$ Scuti\\
$J0904-0532$  & 09:04:15.727 & $-$05:32:47.04 & 45.34  & 19.79 & 19.30 & 19.33 & 19.43 & 19.51 & $\delta$ Scuti\\
              &              &                & 24.21  &       &       &       &       &       &       \\
\hline
\hline
Object & RA & Dec & Period &        u  &         g  &         r  &         i  &         z  & Type\\
       &    &     & (min)  &           &            &            &            &            &\\
$J0859-0429$  & 08:59:27.222 & $-$04:29:16.34 & 7.76   & 20.93 & 20.54 & 20.76 & 20.91 & 21.06 & ZZ Ceti\\
              &              &                & 5.30   &      &      &      &      &      &            \\
$J0900-0442$ \tnote{a}  & 09:00:51.516 & $-$04:42:49.18 & 14.46  & 20.97 & 20.60 & 20.61 & 20.11 & 19.67 & ZZ Ceti\\
                 &              &                & 12.62 &      &      &      &      &      &        \\
\hline
\hline
Object & RA & Dec & Period &        u  &         g  &         r  &         i  &         z  & Type\\
       &    &     & (min)  &           &            &            &            &            &\\
$J0900-0352$  & 09:00:53.225 & $-$03:52:03.96 & 162.90 & 23.51 & 22.30 & 21.67 & 21.41 & 21.28 &  \dots  \\
$J0902-0502$  & 09:02:46.931 & $-$05:02:45.10 & \dots  & 16.98 & 15.89 & 15.33 & 15.16 & 14.71 &  \dots  \\   
$J0904-0516$  & 09:04:02.124 & $-$05:16:45.78 & 454.26 & 21.90 & 22.44 & 22.39 & 22.72 & 22.18 &  \dots  \\
              &              &                & 83.38  &       &       &       &       &       &  \dots  \\
$J0905-0511$  & 09:05:08.181 & $-$05:11:26.75 & 106.67 & 20.24 & 18.43 & 17.58 & 17.21 & 17.04 &  \dots  \\
\hline
\end{tabular}
\begin{tablenotes}
\item[a] \citet{belardi16}
\item[b] \citet{drake13}
\item[c] \citet{drake14}
\end{tablenotes}
\end{threeparttable}
\end{table*}

\subsection{RR Lyrae}

There are three previously known RR Lyrae in our field that were identified as part
of the Catalina survey \citep{drake13,drake14}. Figure \ref{fig:rr} shows the phase-folded
light curves for these objects. They display $\geq 0.5$ mag variations with
periods ranging from 8 to 15 h. We recover, and independently confirm, the pulsation periods
of these three objects in our DECam data. Our phase-coverage for all three pulsators is
excellent, demonstrating that we are able to identify variable stars at hour
and $\sim$day long periods. We do not find any additional RR Lyrae in this field.

\begin{figure}
\includegraphics[width = 9cm]{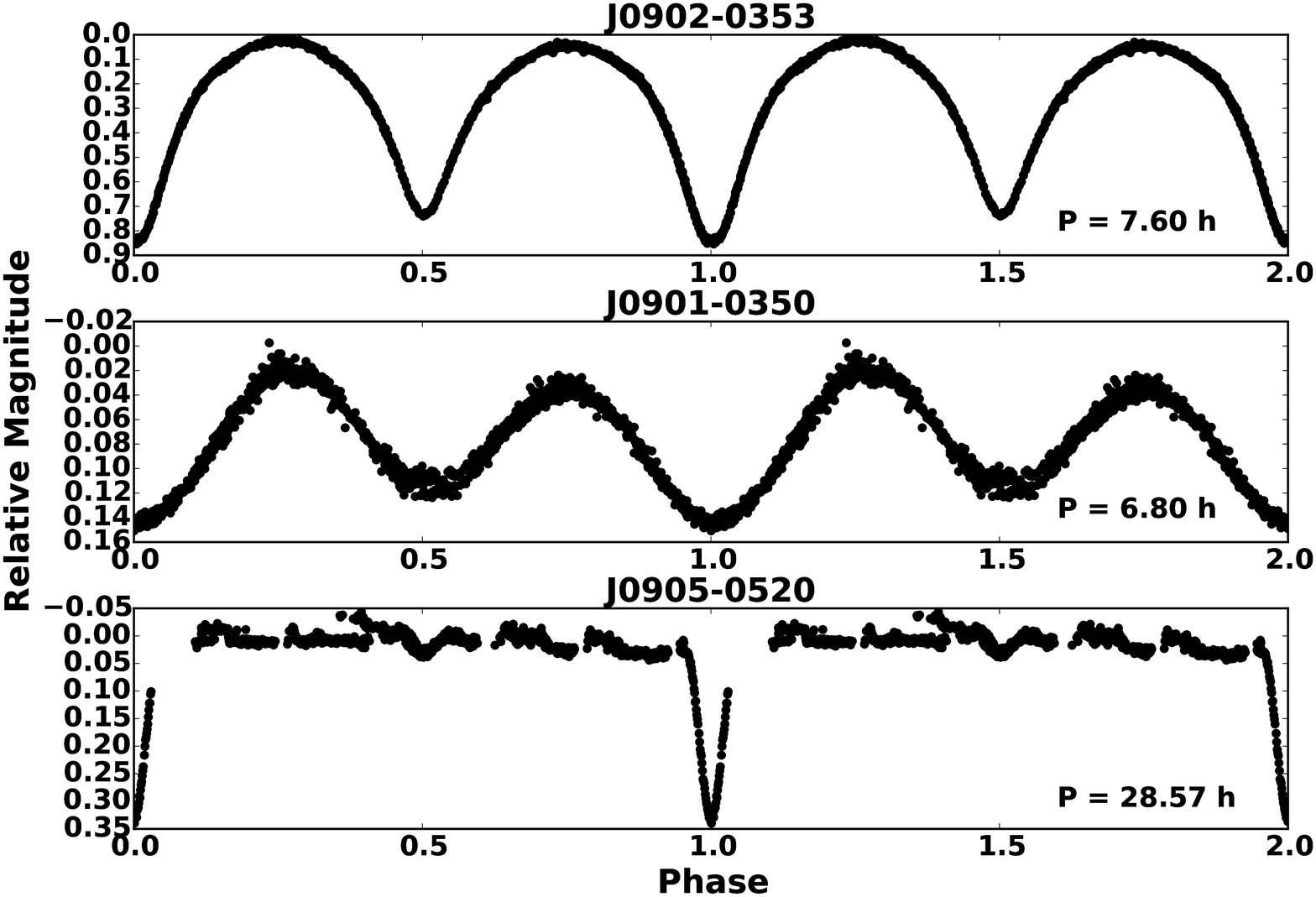}
\caption{Phase-folded light curves for two previously
identified contact binaries (top panels) and a detached binary
(bottom panel) from \citet{drake14}.}
\label{fig:known_binaries}
\end{figure}

\subsection{Contact and Detached Binaries} 
\label{sec:binary} 

There are six previously known contact and detached binaries in our field that
were identified by \citet{drake14}. Figure \ref{fig:known_binaries} shows
the phase-folded light curves for three of these objects. They display $\geq$0.15 mag
variations with binary periods ranging from 6.1 to 28.6 h. Again, our phase coverage
is excellent, and our observations cover $>90$\% of the binary orbit even for the
longest period binary in this sample, J0905$-$0520 with $P=28.6$ h. 

\begin{figure}
\includegraphics[width = 10cm]{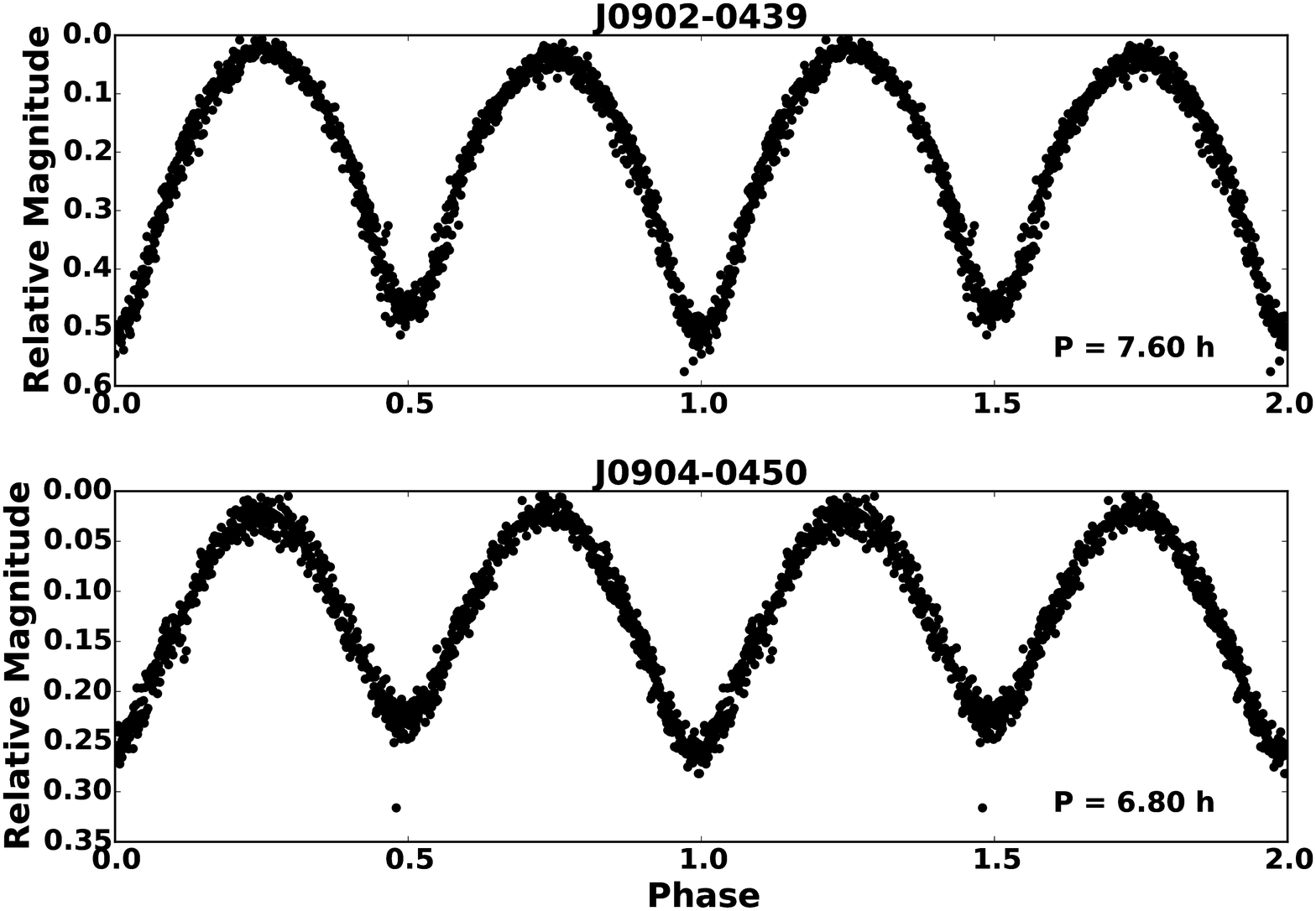}
\includegraphics[width = 10cm]{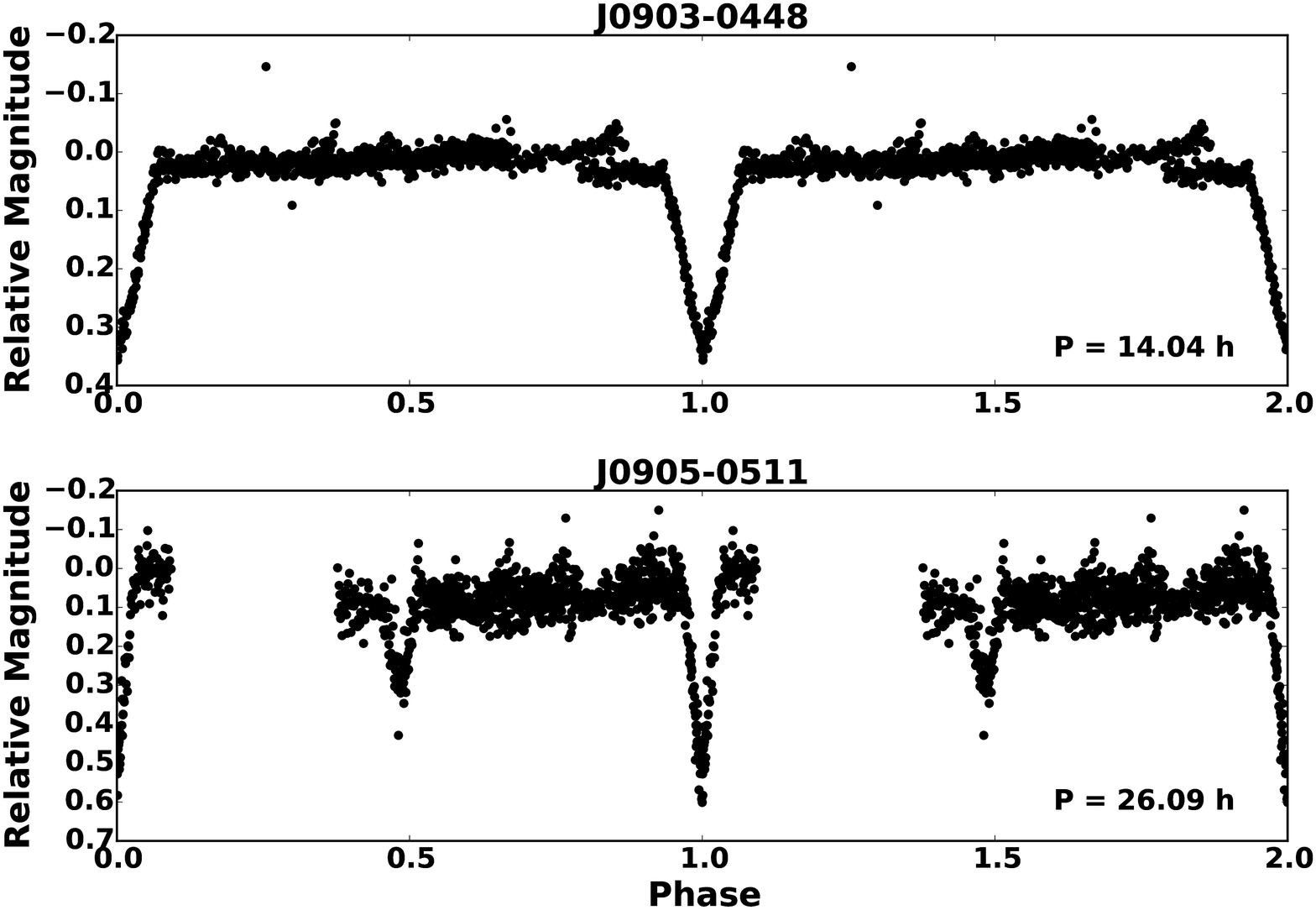}
\caption{Representative phase-folded light curves for four of our new contact (top panels)
and detached (bottom panels) binaries.}
\label{fig:detached}
\end{figure}

We identified an additional set of 18 new binary systems with our DECam data; 7
contact binaries and 11 detached binaries; these are shown in Figure
\ref{fig:colour} as black and yellow triangles respectively. Figure
\ref{fig:detached} shows example phase-folded light curves for four of our
new contact and detached binary systems. The contact binaries have periods ranging
from 6 to 9 hours, while the detached binaries have periods covering a much wider range,
from 6 h to 4 days. However, for five of our 11 detached binaries, the periods
are poorly constrained, and for another three, we could not determine a period
at all. As mentioned above, this is likely due to poor phase coverage for longer
orbital periods. For six of our 11 detached binaries, we observe only one or two
transits over the entire eight half-nights (see Fig. \ref{fig:detached}); in comparison,
our contact binaries have periods short enough to observe a transit every half-night,
allowing for much better constraints on their orbital periods.

\begin{figure}
\includegraphics[width = 10cm]{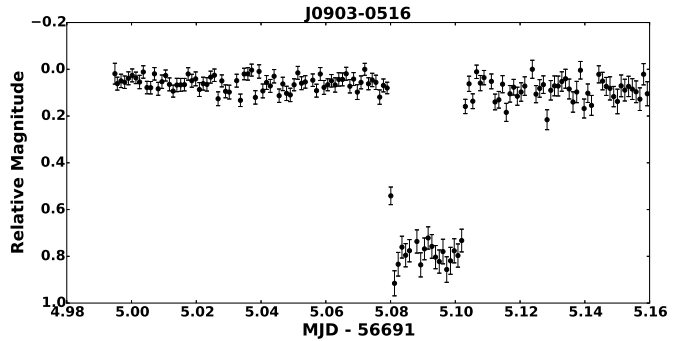}
\caption{The single observed transit for the white dwarf + M dwarf binary J0903-0516.}
\label{fig:wddm}
\end{figure}

One of the new detached binaries involves an eclipsing compact object.
Figure \ref{fig:wddm} shows our light curve for J0903$-$0516. 
We observe only a single transit, so we cannot determine a period.
However, the short ($\approx$ 30 min) duration and the sharp ingress and egress
of the transit mark this object as a white dwarf + M dwarf binary. The observed
light curve is essentially identical to the other eclipsing white dwarf + M dwarf binaries
known \citep[e.g.,][]{parsons17}. About 22\% of field white dwarfs have late-type stellar
companions \citep{farihi05}, while 3.4\% of these systems are eclipsing with $P < 2$ day orbits
\citep{parsons13}. Hence, the frequency of eclipsing white dwarf + M dwarf binaries is likely
around 0.75\% for randomly oriented orbits. Given the predicted number of white dwarfs from the
Besancon Galaxy model (see \S 4), we would expect to find up to 3 eclipsing white dwarf + M dwarfs
in our dataset. Hence, the discovery of such a system in our survey is not surprising.

\subsection{$\delta$ Scuti Type Pulsators}

\begin{figure}
\includegraphics[width = 10cm]{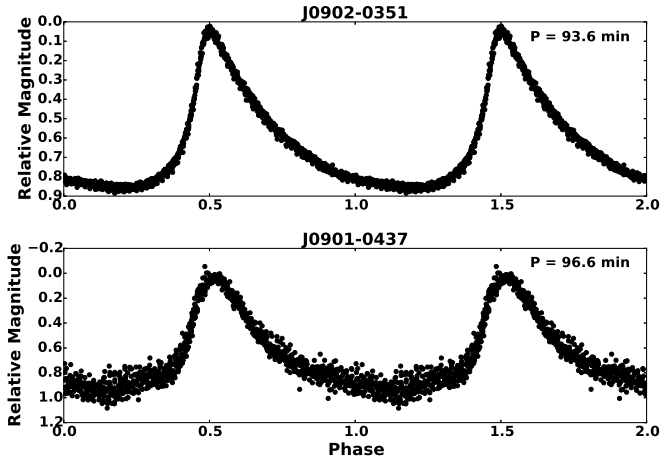}
\caption{Phase-folded light curves for two of the newly identified $\delta$ Scuti type pulsators.}
 \label{fig:delta}
\end{figure}

\begin{figure}
\includegraphics[width = 10cm]{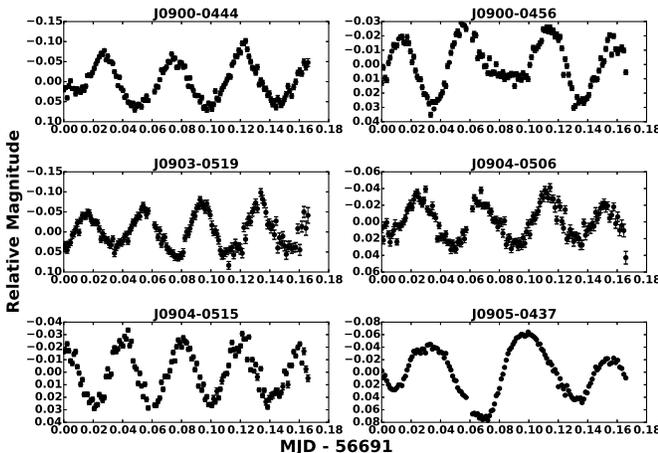}
\caption{Light curves from a single night for 6 additional $\delta$ Scuti type pulsators in our DECam field.}
\label{fig:del}
\end{figure}

$\delta$ Scuti stars are the second most common type of variable stars in our survey. They
typically have A-F spectral types and pulsation periods of 0.48-6 hrs
\citep{chang13}. We find 15 objects with colours and periods consistent with
$\delta$ Scuti type pulsators. These objects are shown as blue diamonds in
Figure \ref{fig:colour}. We show the phase-folded light curves for two of the newly identified
$\delta$ Scuti type pulsators with periods near 1.6 h in Figure \ref{fig:delta}, and
the light curves for a sample of the remaining from a single night
in Figure \ref{fig:del}. 

All but one of these pulsators have periods ranging from 55 min
to 97 min. However, there is a relatively red object, J0906-0407 with $g-r=0.52$ mag,
that shows variations with a dominant period of 15 min, which is near the period minimum
of $\approx$18 min observed in the Kepler mission $\delta$ Scuti star sample \citep{uytter11}.

\begin{figure}
\includegraphics[width = 10cm]{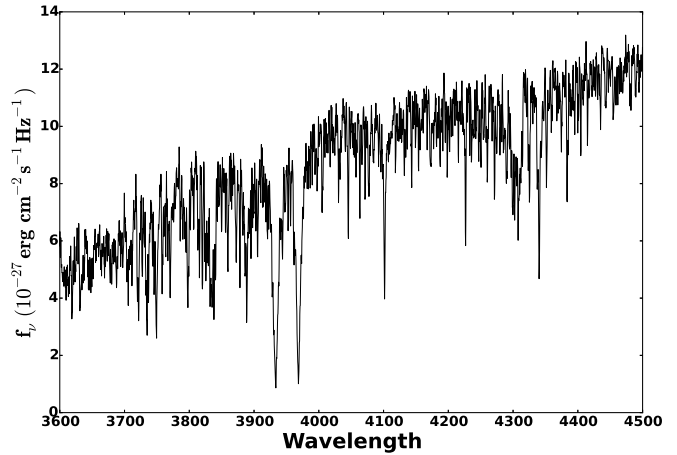}
\caption{MMT spectrum of J0906-0407, a newly identified G-type $\delta$ Scuti pulsator.}
\label{fig:j0906}
\end{figure}

Given the relatively short period that could either indicate a pulsating ZZ Ceti white dwarf or
a short period $\delta$ Scuti, we obtained follow-up spectroscopy of J0906-0407 to constrain its nature.
Figure \ref{fig:j0906} shows its optical spectrum, which reveals strong Ca II H \& K lines and the G-band absorption,
as well as relatively weak H$\delta$ and H$\gamma$ lines. Hence, J0906-0407 is clearly a G star.
We compare the equivalent width measurements of these features to the spectral templates from
\citet{pickles98} and identify this object as a $\approx$G0V star.
\citet{macfarlane17} identify two $\delta$ Scuti pulsators with dominant periods
of 9.0-9.6 min, including the G5 type star OW J075531.59-315058.2. Hence, J0906-0407 appears to be similar to other 
G-type $\delta$ Scuti stars with relatively short pulsation periods.

\subsection{Short Period Pulsators} 

\begin{figure*}
\includegraphics[width = 17cm]{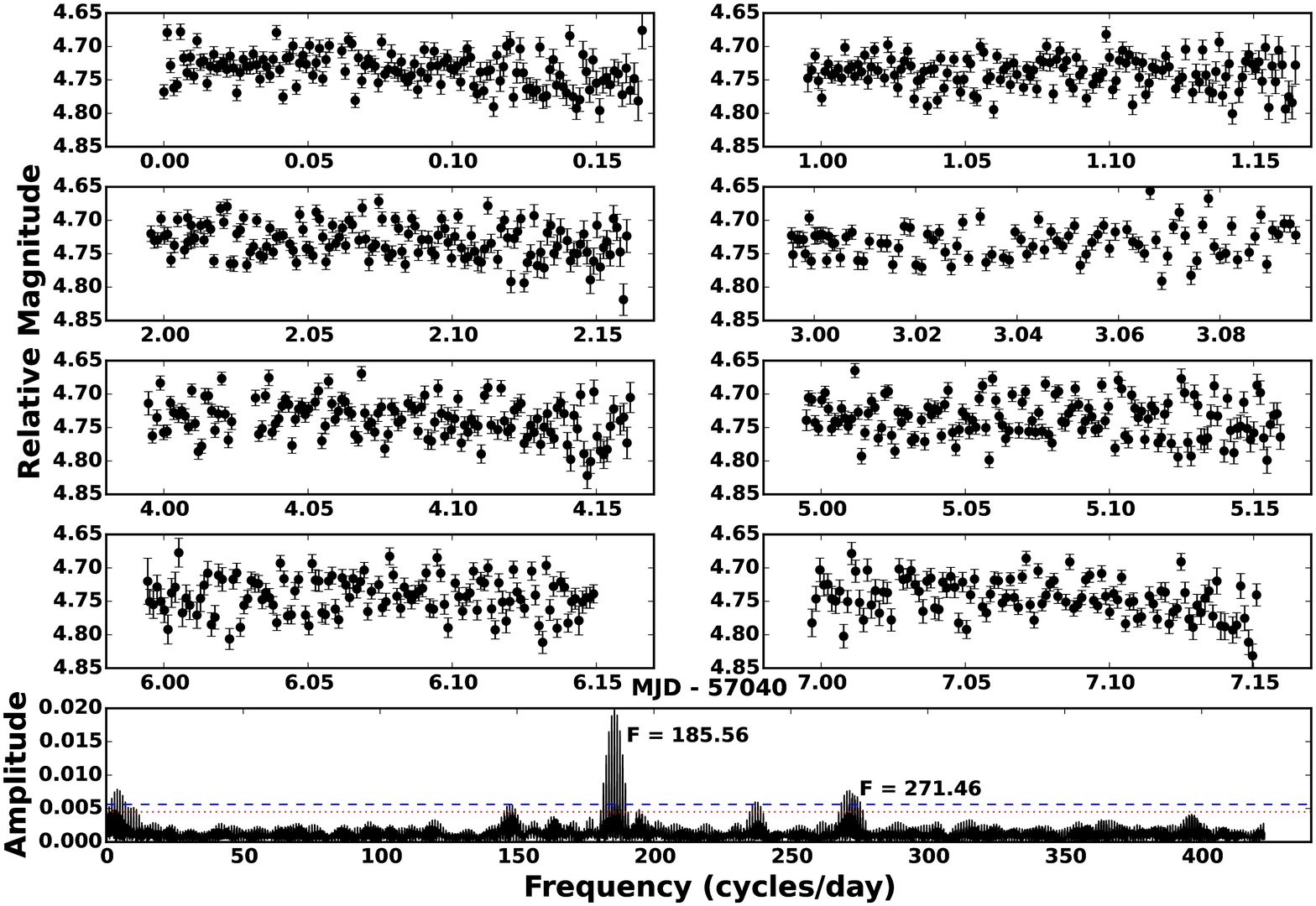}
\includegraphics[width = 17cm]{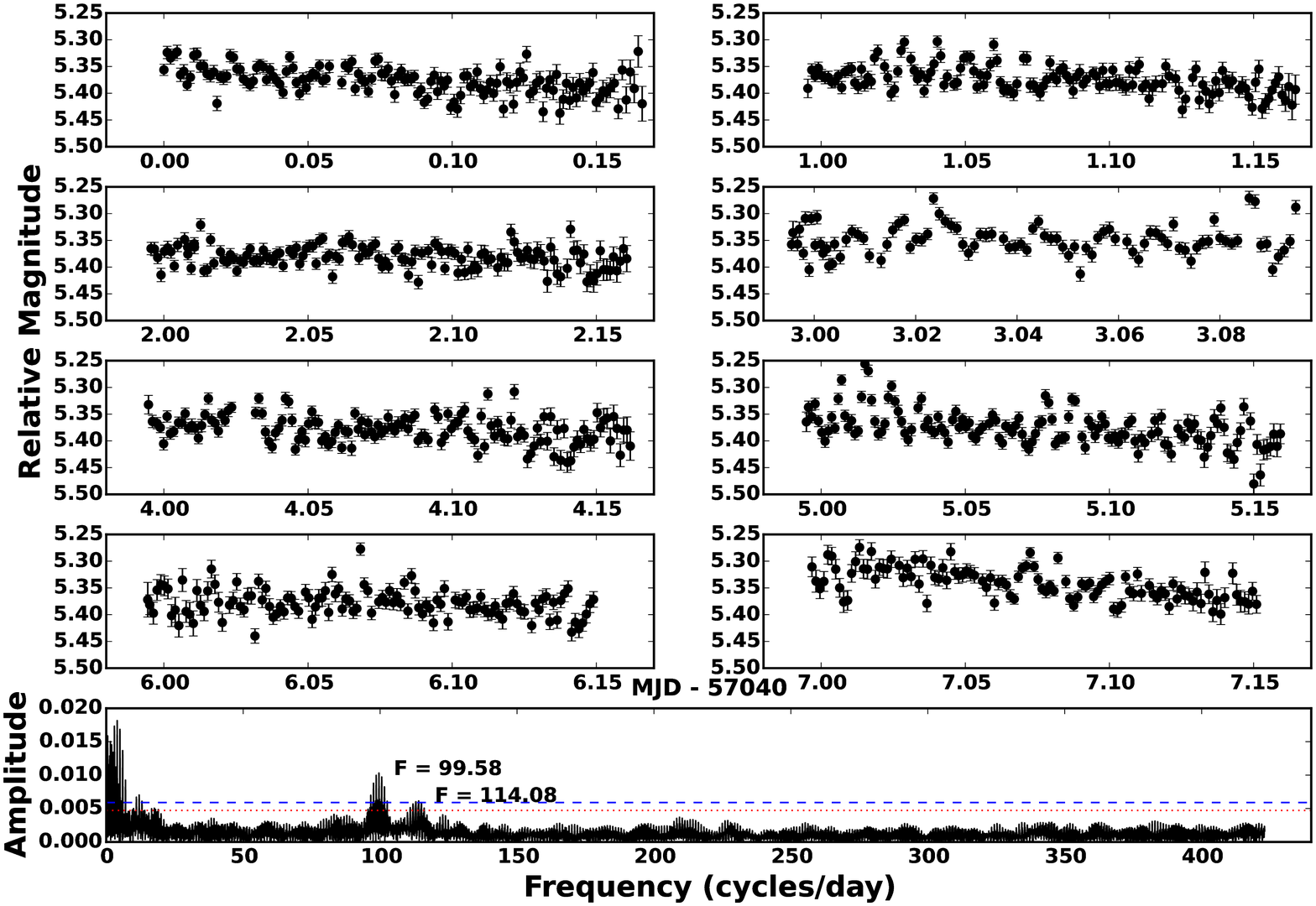}
\caption{Light curves (top panels) and Fourier transforms
(bottom panel) for J0859-0429 (top) and J0900-0442 (bottom). The red dotted
line and blue dashed line in the Fourier transform show, respectively, the
4$\langle$A$\rangle$ and 5$\langle$A$\rangle$ detection limits, where 
$\langle$A$\rangle$ the median amplitude of the Fourier transform.}
\label{fig:zzceti}
\end{figure*}

Among the 7 objects that cannot be readily
identified by colour and period, we find three blue objects with short periods
(P $<$ 60 min). Their location in the $u-g$ versus $g-r$ color-color diagram (see
Figure \ref{fig:colour}) is consistent with DA white dwarfs.
Figure \ref{fig:zzceti} shows the light curves and their Fourier transforms
for two of these objects. The first is J0900-0442, the ZZ Ceti candidate previously 
discovered in \citet{belardi16} through its high proper motion.
We detect two significant periods at 14.46 and 12.62 min, consistent with those found
previously. Additionally, we find a second ZZ Ceti candidate in the field. J0859-0429
exhibits significant variations at 7.76 and 5.30 min, consistent with pulsation periods
observed in other ZZ Ceti white dwarfs \citep{fontaine08,winget08}.

J0859-0429 has $ugriz$ photometry that puts it right in the middle of the ZZ Ceti instability 
strip for white dwarfs \citep{gianninas11} and J0900-0442 displays a proper motion of 27 mas 
yr$^{-1}$ which puts it right in the middle of the white dwarf sequence in the reduced proper 
motion diagram \citep{belardi16}. As discussed in \citet{belardi16}, the inability of the models 
to reproduce the SED of J0900-0442 may be due to an unseen companion. The remaining blue object 
that shows short period pulsations (at 24.21 and 45.34 min), J0904-0532, has colours consistent 
with a white dwarf, and its spectral energy distribution is well fit by a $\approx$8600 K object. 
These pulsation periods are longer than expected and the temperature is cooler than expected for 
average mass ZZ Ceti white dwarfs.

\begin{figure}
\includegraphics[width = 10cm]{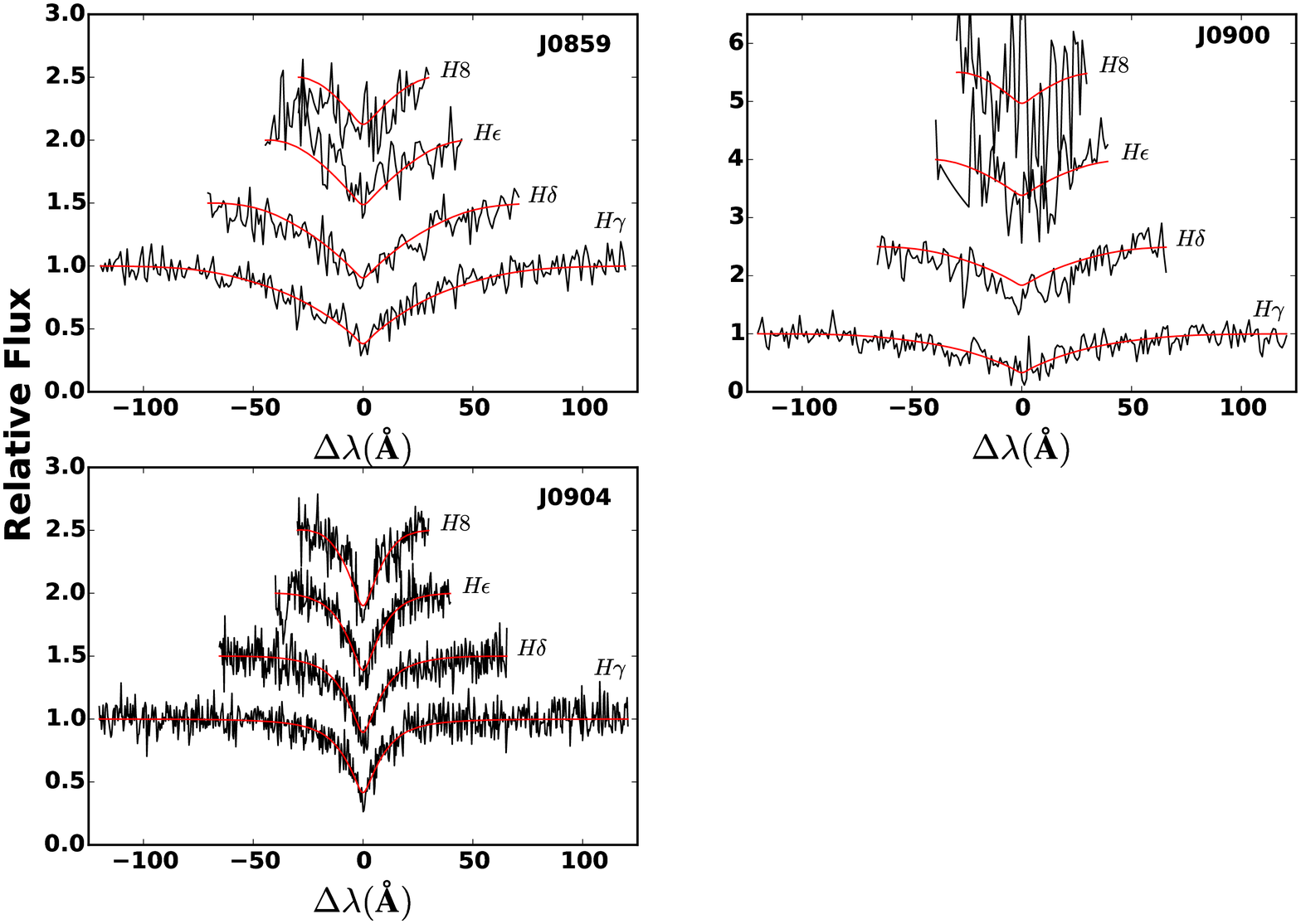}
\caption{1D model fits (red) to the observed Balmer line profiles (black) for J0859-0429 (top left), 
J0900-0442 (top right), and J0904-0532 (bottom left). For clarity, line profiles are vertically offset
from each other.}
\label{fig:spectra}
\end{figure}

We obtained follow-up optical spectra of J0859-0429 and J0900-0442 from Gemini South, and
J0904-0532 from MMT. The observed line profiles(black) and 1D model fits \citep[red,][]{gianninas11} are 
shown in Figure \ref{fig:spectra}. 

The normalised Balmer line profiles of J0859-0429 are best-fit by a pure H atmosphere model with 
$T_{\rm eff} = 12940 \pm 690$ K and $\log{g} = 7.94 \pm 0.15$. These values are consistent with the 
instability strip for white dwarfs, confirming that J0859-0429 is a ZZ Ceti. Inclusion of H$\beta$ in 
the fit yields $T_{\rm eff} = 12400 \pm 410$ K and $\log{g} = 7.82 \pm 0.15$; the results are consistent 
within 1$\sigma$.

The normalised Balmer line profiles of J0900-0442 are best-fit by a pure H atmosphere model with 
$T_{\rm eff} \sim 12140$ K and $\log{g} \sim 6.85$. These values are broadly consistent with 
the instability strip; however, the errors are too large to unequivocally confirm that J0900-0442 is a ZZ Ceti.
Since the SED of J0900-0442 shows contamination from a late type star, we do not include H$\beta$ in the fit.

The normalised Balmer line profiles of J0904-0532 are best-fit by a pure 
H atmosphere model with $T_{\rm eff} = 8350 \pm 80$ K and $\log{g} = 5.46 \pm 0.16$.
Fitting synthetic main-sequence star spectra with pure H atmosphere white dwarf models,
\citet{brown17} demonstrate that these models systematically overestimate the surface gravity
below 9000 K \citep{pelisoli18}. For example, the synthetic spectrum of a main-sequence star with [Fe/H] = $-1$,
$T_{\rm eff}= 8000$ K, and $\log{g}=4.5$ is best-fit by a white dwarf model with $\log{g}\approx 5.1$.
Hence, the model atmosphere fit is likely an overestimate of the actual surface gravity of this star. Additionally,
its position in Figure \ref{fig:cmd} places it in the middle of the 15 $\delta$ Scuti type pulsators. Hence, 
J0904-0532 is certainly a $\delta$ Scuti type variable star.

\subsection{Variables with Undetermined Types}

There remains four objects for which we cannot determine the source of variability
from colour and period alone. These objects are listed at the bottom of Table 1 and
are marked by green circles in Figure \ref{fig:colour}.
J0900-0352, J0902-0502, and J0905-0511 have colours consistent with main-sequence stars,
but the light curves are simply too inconclusive to
make a definitive identification. J0904-0516 has colours consistent with
quasars; however, it exhibits variations with periods of 454.26 and 83.38 min, far shorter than
the typical periods previously observed in quasars. Follow-up spectroscopy would be
helpful in revealing the nature of these four objects, though two of them are relatively faint
with $g>22$ mag.

\section{Conclusions} \label{sec:con}

We present final results from the first field of the DECam minute-cadence
survey. We construct light curves for the 31732 point source in the field,
allowing us to search for potential planetary transits around otherwise
difficult to identify low proper motion white dwarfs ($\mu$ $<$ 20 mas
$yr^{-1}$). While we find no compelling evidence of such a transit around a
white dwarf, we do identify 40 new variable objects: 18 binary systems
(including a white dwarf + M dwarf binary), 16 $\delta$ Scuti type pulsators, and 
two ZZ Ceti white dwarfs. The remaining variable objects of undetermined type will 
require follow-up spectroscopic observations for reliable classification.

Analysis of the second and third fields is underway and will be presented in a future
publication. We expect similiar numbers of white dwarfs in the remaining two
fields, allowing us to place stringent constraints on the frequency of Earth-sized planets
in the habitable zone of white dwarfs. Additionally, surveys like SuperWASP,
the Next-Generation Transit Survey, the Transiting Exoplanet Survey Satellite, 
and ZTF will obtain high-cadence observations for millions of stars. These observations 
will allow us to probe the variability of the night sky on timescales that remain 
largely unstudied, and open the possibility of observing previously unknown phenomena.

\section*{Acknowledgements}

We thank the referee for their helpful suggestions. This work is in part supported by NASA under grant NNX14AF65G.
Based on observations at Cerro Tololo Inter-American Observatory, National Optical
Astronomy Observatory (NOAO Prop. ID:2014A-0073 and PI: M. Kilic), which is operated by the Association
of Universities for Research in Astronomy (AURA) under a cooperative agreement with
the National Science Foundation. 

This project used data obtained with the Dark Energy Camera (DECam), which was constructed
by the Dark Energy Survey (DES) collaboration. Funding for the DES Projects has been provided by
the DOE and NSF (USA), MISE (Spain), STFC (UK), HEFCE (UK). NCSA (UIUC), KICP (U. Chicago), CCAPP
(Ohio State), MIFPA (Texas A\&M), CNPQ, FAPERJ, FINEP (Brazil), MINECO (Spain), DFG (Germany) and
the collaborating institutions in the Dark Energy Survey, which are Argonne Lab, UC Santa Cruz, 
University of Cambridge, CIEMAT-Madrid, University of Chicago, University College London, 
DES-Brazil Consortium, University of Edinburgh, ETH Zurich, Fermilab, University of Illinois, ICE
(IEEC-CSIC), IFAE Barcelona, Lawrence Berkeley Lab, LMU Munchen and the associated Excellence
Cluster Universe, University of Michigan, NOAO, University of Nottingham, Ohio State University,
University of Pennsylvania, University of Portsmouth, SLAC National Lab, Stanford University,
University of Sussex, and Texas A\&M University.

\end{document}